\input harvmac
\def\np#1#2#3{Nucl. Phys. B{#1} (#2) #3}
\def\pl#1#2#3{Phys. Lett. {#1}B (#2) #3}

\def\physrev#1#2#3{Phys. Rev. {D#1} (#2) #3}

\def\ev#1{\langle#1\rangle}

\def\tilde{\widetilde}
\def\frac#1#2{{#1\over#2}}
\def\half{\frac{1}{2}}
\def\al{\alpha}

\def\Pf{{\rm Pf}}

\Title{hep-th/9602031, RU-95-78}
{\vbox{\centerline{Duality and Dynamical Supersymmetry Breaking}
\centerline{}\centerline{in $Spin(10)$ with a Spinor}}}
\bigskip
\centerline{P. Pouliot and M.J. Strassler\footnote{*}{{e-mail 
address: pouliot, strasslr@physics.rutgers.edu}}}
\vglue .5cm
\centerline{Department of Physics and Astronomy}
\centerline{Rutgers University}
\centerline{Piscataway, NJ 08855-0849, USA}

\bigskip

\noindent

We study ${\cal N}=1$ supersymmetric $Spin(10)$ chiral gauge theories 
with a single spinor representation and $N$ vector representations.
We present a dual description in terms of an ${\cal N}=1$ 
supersymmetric $SU(N-5)$
chiral gauge theory with a symmetric tensor, 
one fundamental and $N$ antifundamental representations. 
The $Spin(10)$ theory with $N=0$ breaks supersymmetry at strong coupling;
we study how this arises at weak coupling in the dual theory, which 
is a spontaneously broken gauge theory.  
Also, we recover various known dualities, 
find new dual pairs and generate
new examples of dynamical supersymmetry breaking.

\Date{1/96}

\nref\witt{E. Witten, \np{185}{1981}{513}}
\nref\adsdsb{I. Affleck, M. Dine and N. Seiberg, \np{256}{1985}{557}}
\nref\adsspin{I. Affleck, M. Dine and N. Seiberg, \pl{140}{1984}{59}}
\nref\nelson{A.E. Nelson and N. Seiberg, hep-ph/9309299,
\np{416}{1994}{46}}
\nref\iss{K. Intriligator, N. Seiberg and S. Shenker, hep-ph/9410203,
\pl{342}{1995}{152}}
\nref\pop{E. Poppitz and S. P. Trivedi, hep-th/9507169,
EFI-95-44}
\nref\dine{M. Dine, A.E. Nelson, Y. Nir and
Y. Shirman, SCIPP 95/32, UW-PT/95-08, WIS-95/29/Jul-PH,  hep-ph/9507378}
\nref\kinsrev{K. Intriligator and N. Seiberg, 
hep-th/9509066, RU-95-48, IASSNS-HEP-95/70}
\nref\sem{N. Seiberg, hep-th/9411149, \np{435}{1995}{129}}%
\nref\ithomas{K. Intriligator and S. Thomas,
SLAC-PUB-95-7041, to appear}
\nref\suantisym{P. Pouliot, hep-th/9510148, RU-95-66}
\nref\mur{H. Murayama,  hep-th/9505082, \pl{355}{1995}{187}}
\nref\son{K. Intriligator and N. Seiberg,  hep-th/9503179,
\np{444}{1995}{125}; 
hep-th/9506084, RU-95-40, IASSNS-HEP-95/48}
\nref\kutschsei{D. Kutasov, A. Schwimmer and N. Seiberg,
 hep-th/9510222,  EFI-95-68, WIS/95/27, RU-95-75}
\nref\spinseven{P. Pouliot, hep-th/9507018,  \pl{359}{1995}{108}}
\nref\spineight{P. Pouliot and M. Strassler, hep-th/9510228,  RU-95-67}
\nref\nativacua{N. Seiberg, hep-th/9402044,
\physrev{49}{1994}{6857}}
\nref\ilstr{K. Intriligator, R.G. Leigh and M.J. Strassler,
hep-th/9506148, RU-95-38}

\newsec{Introduction}

For supersymmetry to be relevant to the real world, it must be broken 
somehow. Dynamical supersymmetry breaking \witt\ is one interesting 
possibility. Until recently, few models with this feature had been 
found \refs{\adsdsb, \adsspin}, but
now a great many more are known
\refs{\nelson - \dine}. 
In the last two years, there has been considerable progress
in the understanding of the low-energy dynamics of supersymmetric 
field theories \kinsrev.  The unifying principle
in this endeavor is duality \sem: two 
different gauge theories may have
exactly the same long distance physics. 
But to establish duality, supersymmetry is essential.

In this letter, as well as in \refs{\ithomas, \suantisym}, 
the dynamical breaking of supersymmetry 
is studied in a theory which exhibits duality.
We consider an ${\cal N}=1$ supersymmetric $Spin(10)$ 
gauge theory. Its matter content consists of $N$ 
superfields $Q$
in the ${\bf 10}$-dimensional (vector) representation 
and a field $P$ in  
the ${\bf 16}$-dimensional (spinor)\foot{$SO(10)$ and $Spin(10)$
have the same algebra, but only the latter has spinor representations.
$Spin(10)$ has two inequivalent complex spinor representations ${\bf 16}$
and ${\bf \overline{16}}$; while ${\bf 16}\otimes{\bf \overline{16}}$
contains a singlet, ${\bf 16}\otimes{\bf 16}$ does not.} representation
of the gauge group and it has a tree-level superpotential
$W=0$.

When $N=0$, the classical scalar potential
has no flat direction. It was shown in \adsspin\ that supersymmetry is 
broken dynamically on the basis that no plausible low-energy description
could be found satisfying the 't Hooft anomaly matching conditions.
In \mur, the $N=1$ theory was used to prove that
no vacuum state exists at weak coupling, consistent with
the result of \adsspin. In this letter, we present new and
independent evidence that no vacuum exists at strong coupling either.

To this end, we will study a dual for these $Spin(10)$ theories.
The dual ``magnetic'' theory we propose
is, for $N\ge 7$, an ${\cal N}=1$ supersymmetric gauge theory
with gauge group $SU(N-5)$, with 
$N$ matter superfields $q$ in the antifundamental representation, a 
field $q'$ in the fundamental representation, and a field $s$ in the
symmetric tensor representation.  There are also 
elementary gauge singlets $M$ and 
$Y$ in the dual, to be identified by duality 
with composite gauge singlets $M=Q^2$ and $Y=QP^2$ of the $Spin(10)$ 
theory. These fields interact via a superpotential 
$W_{magnetic}=\det s/\mu_1^{N-8} +Mqsq/\mu_2^2
+Yqq'/\mu_3^2$.\foot{As emphasized in \refs{\sem, \son, \kutschsei}, the
scales $\mu$ are crucial for the details of the duality; we will,
however, set $\mu=1$, 
as our purposes do not require keeping track of them.}
When $N=7$, an extra invariant term $M^6Y^2$ must be added to
$W_{magnetic}$ in order to reproduce known results.
The duality presented here is an extension of the one 
found in \spinseven\ for $G_2$ and $Spin(7)$
(see also \spineight\ for an intermediate step to $Spin(8)$.)

\newsec{Checks of the Duality}
 
The first step to analyze the physics is to find the independent 
gauge-invariant chiral operators. 
In the ``electric'' $Spin(10)$ theory, 
there are mesons $M_{ij}=Q_i Q_j$  which are symmetric in their
flavor indices $i,j=1,\ldots, N$, 
and $Y_{i}=Q_i PP$. $M$ and $Y$ are
also present in the magnetic theory, but as elementary fields.  
There are also a 
number of baryons or exotics, totally antisymmetric in flavor,
namely $B=Q^5P^2$, 
$D_n=Q^{6+2n} W_\al^{2-n}$, and
$E_n=Q^{5+2n}P^2 W_\al^{2-n}$, $(n=0,1,2)$, where $W_\al$ is
the gauge superfield strength.\foot{In $B$ the two $P$ fields
are combined in a ${\bf 126}$ representation (a self-dual
five-vector-index tensor) while in $E_n$ the two $P$ fields
are combined in a vector representation; all of the vector indices
are then contracted with a ten-index epsilon tensor.}  
In the magnetic theory, there are baryons
$\tilde B\equiv q^{N-5}$ contracted with one (N-5)-index epsilon 
tensor, $\tilde D_n=q^{N-6-2n}s^{N-6-n}q' \tilde W^n_\al$
and $\tilde E_n=q^{N-5-2n}s^{N-5-n} \tilde W^n_\al \ (n=0,1,2)$,
contracted with two (N-5)-index epsilon tensors. 
The operators $\det s$, $q's^{N-6}q'$,
$qsq$, $qq'$ are redundant because of the
equations of motion following {}from the superpotential.
 
There are many consistency checks on this duality.
First, as required by duality, the global symmetry of the
magnetic theory is the same as
that of the electric theory,
namely
$SU(N)\times U(1)\times U(1)_R$
(there is no discrete symmetry).
 Under this symmetry
 the matter fields $Q$ and $P$ of the electric theory transform as
$({\bf N},\ -1,\ 1-{8\over N+2})$ and
$({\bf 1},\ \half N,\ 1-{8\over N+2})$
while the matter fields of the magnetic theory 
$s$, $q$, $q'$ transform as
$({\bf 1},\ 0,\ {2\over N-5})$,  
$({\bf\overline N},\ 1,\ {8\over N+2}-{1\over N-5})$ and
$({\bf 1}, -N, -1+{16\over N+2}+{1\over N-5})$.
Note that our choice of a basis for the $R$-symmetry $U(1)_R$ is
arbitrary.
A highly non-trivial check is
that the 't Hooft anomaly matching conditions are satisfied
at the origin of the moduli space of vacua.  
The 
symmetries, holomorphy and weak coupling 
forbid any dynamically generated 
superpotential for  $N\ge 6$, as 
is clear {}from the above choice of $R$-charges
(we use the usual convention that the superpotential has $R$-charge 2.)
The symmetries, holomorphy, and smoothness near
the origin $M, Y, s, q, q'=0$ uniquely determine the
magnetic superpotential for $N> 7$. 

There is a one-to-one correspondence, preserving all global
symmetries, 
between the gauge-invariant operators $(M, Y, B,
 D_n,E_n)$
of the electric theory and the operators $(M, Y, \tilde B, \tilde D_n,
\tilde E_n)$ of
the magnetic theory.
Some of these operators are constrained.
  In a similar duality mapping \spineight\
we have shown that the mapping of the constraints is consistent,
using the chiral anomaly and non-perturbative dynamics.
However, we did not perform such checks in detail for the 
operators here.

For $N\ge 22$, the electric theory is not asymptotically
free, so it flows to a free theory of $Spin(10)$ gluons and quarks
in the infrared, as does the magnetic theory. 
For $7 \leq N \leq 21$, we expect, but cannot prove, that  
the electric and magnetic theories are in a non-trivial
interacting superconformal phase at the origin.

To check the duality further, we consider certain deformations.
If we give an expectation value $\vev{Y_N}\neq 0$
to a component of $Y$, the $Spin(10)$ group is broken to $Spin(7)$
by the Higgs mechanism.
The fields $Q_N, P$ are eaten by the massive gauge bosons and 
$N-1$ eight-dimensional spinors $\hat Q$
of $Spin(7)$ remain massless. 
The other two components of $Q_i$, $i=1,\ldots, N-1$,
 become decoupled
singlet fields.
In the magnetic theory, the expectation value for $Y_N$
gives mass to $q^N$ and $q'$, leaving an $SU(N-5)$ theory with
$N-1$ fields $q$, a symmetric tensor $s$, 
a superpotential $W=\det s+Mqsq$,
and decoupled singlets $M_{iN}$ and $Y_i$; this is the duality
expected {}from \spinseven. When $N=7$, 
the extra term $M^6Y^2$ in $W_{magnetic}$
becomes $\det \hat M$ \spinseven, where $\hat M=\hat Q^2$. 
The dynamical origin of the terms
$M^6Y^2$ and $\det \hat M$, which carry two-instanton quantum numbers,
 is not presently understood.
 
\newsec{More dualities}

A set of new dualities is found if we go along flat directions
with $\vev{M}$ having rank $r$.  In this case, the electric theory
breaks to $Spin(10-r)$, with spinors in a sixteen-dimensional
reducible representation and $N-r$ vectors.  In the magnetic theory,
the superpotential becomes $W=\det s + \left[\vev{M}+M\right]qsq +Yqq'$.
One direction is to explore these new theories in detail. Instead, for
conciseness, 
we will mention only two of
them here, and show that they match to previously known dualities.

Consider the
case $r=1$. For convenience, 
let us take $N+1$ vectors $Q_i$, $i=0,1,\ldots, N$
in the electric $Spin(10)$ theory. Say $\ev{M_{00}}\neq 0$. 
The resulting $Spin(9)$ theory has $N$ vectors and
a sixteen-component spinor $\hat P$.
While no mass term for $P$ could be written in $Spin(10)$,
a mass term exists for $\hat P$ in $Spin(9)$, namely $Y_0 = \vev{Q_0}P^2$ 
in terms of $Spin(10)$ fields. When we add $Y_0$ to the superpotential
of $Spin(9)$, the
electric theory becomes $SO(9)$ with $N$ vector representations $\hat Q$.
 When we add $Y_0$ to the magnetic superpotential
$W=\det s + \ev{M_{00}}q^0sq^0 + Mqsq + Yqq' + Y_0$,
the operator $q^0q'$ condenses.
This breaks the dual $SU(N-4)$ gauge group 
to $SU(N-5)$.  {}From the equation of motion for $s$, we have
$s^{-1}\det s  + M q q = - \ev{M_{00}} q^0q^0$. Since $M$ is kept
arbitrary in the electric theory, this implies $s^{-1}\det s \neq 0$.
The gauge group is thus broken further to $SO(N-5)$.
It can be checked that, after all the massive
fields are integrated out, the remaining fields are
 $N$ vectors $\hat q^i$ of $SO(N-5)$ with the singlets $\hat M_{ij}$,
$i,j=1,\ldots, N$, some decoupled singlet fields
and the superpotential $\hat M\hat q\hat q$, 
as expected {}from the duality between $SO(9)$ and $SO(N-5)$
with $N$ vectors \refs{\sem,\son}.

For our second example,
 we will derive one more new dual pair and use it to
relate the $Spin(10)$ duality to the $Spin(8)$ duality of \spineight.
We start with a similar set-up as in the previous paragraph:
the $Spin(10)$ theory with
$N+1$ vectors $Q$.  First, add
$Y_0$ to the superpotential;
then go along the flat direction $\ev{M_{0N}}\neq 0$.
In the electric theory,
$Spin(10)$ is broken to $Spin(8)$. The ${\bf 16}$-spinor $P$ splits
into an ${\bf 8_s}$-spinor $\hat P$ and an ${\bf 8_c}$-conjugate-spinor.
 Note that under this breaking,
the superpotential $Y_0$ gives a mass only to the ${\bf 8_c}$;
the mass term for $\hat P$ is $Y_N$. The result is $Spin(8)$ with 
$N-1$ vectors $\hat Q$, a spinor $\hat P$ and 
$W_{electric}=U\hat P\hat P$, where $U$ is
the $Spin(8)$-singlet component of $Q_0$ which neither gets an 
expectation value nor is eaten by massive gauge bosons.
In the magnetic $SU(N-4)$ theory,
$W= \det s + Mqsq + \ev{M_{0N}}q^0sq^N+Yqq'+Y_0$.
As above, $SU(N-4)$ is first higgsed to $SU(N-5)$ by $\ev{q^0q'}\neq 0$. 
{}From the term $\ev{M_{0N}}\ev{q^0}sq^N$, the components of $s$ which
are a singlet or a fundamental under $SU(N-5)$ are
massive. Consequently, the $\det s$ term is eliminated.
The result of integrating out the massive fields
is $W= \hat M_{ij} \hat q^i_\al\hat s^{\al\beta}\hat q^j_\al$, 
$i,j=1,\ldots, N-1$, $\al,\beta=1,\ldots, N-5$, 
as expected {}from \spineight.

We note that by adding singlets to both the electric and 
the magnetic theories,
the duality for $Spin(10)$ as an electric theory is
trivially converted into one for $SU(N-5)$ as the electric theory with 
the same charged matter content as the above magnetic theory but
no singlet fields.  Specifically, the dual to $SU(N-5)$ with fields
$s$, $q$, $q'$ and superpotential $W=\det s$ is $Spin(10)$
with fields $Q$, $P$, singlets ${\tilde M}=qsq$, ${\tilde Y}=qq'$, and
a superpotential $W={\tilde M}Q^2+{\tilde Y}QP^2$.

\nref\switten{N. Seiberg and E. Witten, hep-th/9407087,
\np{426}{1994}{19};
hep-th/9408099, \np{431}{1994}{484}}
\nref\olive{C. Montonen and D. Olive, \pl{72}{1977}{117}}

We now show how to flow to a duality in ${\cal N}=2$
supersymmetric $SU(2)$ gauge theory found in \switten.
Consider for the $Spin(10)$ gauge theory 
the case $N=8$. Add to this theory seven gauge singlet fields
$\tilde M^{8j}$, $j=1,\ldots,7$. Take the tree level superpotential to be
$W_{electric}=Q_8P^2+\tilde M^{8j} Q_8Q_j$. Go along a flat direction
with $\ev{M_{ij}}\propto \delta_{ij}$, $i,j=1,\ldots,7$.
The result after integrating out the massive 
fields is a $Spin(3)\cong SU(2)$
gauge theory with a triplet $\hat Q_8$ and eight doublets
$\hat P$, and with the superpotential $\hat Q_8 \hat P \hat P$ that 
renders the theory ${\cal N}=2$ supersymmetric (and 
ultraviolet finite). Now let us study the
dual. It is an $SU(3)$ gauge theory with the
superpotential $W=Mqsq+Yqq'+\det s+ Y_8 + \tilde M^{8j} M_{8j}$,
with $\det s = s^3$.
First $SU(3)$ is broken to $SU(2)$ by $\ev{q^8q'}$.
The field $s$ decomposes into a triplet $\hat s$, a doublet $q^0$ and
a singlet; the fields $M_{88}$, $Y$, $M_{8j}$, $\tilde M^{8j}$
and the $SU(2)$-singlet components of the charged fields are all 
massive and may be integrated out.
Then, with $\ev{M_{ij}}\propto \delta_{ij}$, 
$i,j=1,\dots,7$, the result is simply the superpotential 
$W= \hat s \hat q^0\hat q^0+\hat s\sum \hat q^i\hat q^i$.
This magnetic theory also has ${\cal N}=2$ supersymmetry
and is isomorphic
to the electric one \switten, as expected {}from Montonen-Olive 
duality \olive. 

\newsec{Mass perturbations of the $Spin(10)$ theory}

We now add to the $Spin(10)$ theory a mass term $W=mQQ=mM$, 
where $m$ has rank $r$. We integrate out the massive 
vectors and flow to a $Spin(10)$ theory with $N-r$
vectors.  In the
dual, the operator $qsq$ gets an expectation value of rank $r$,
breaking the color group {}from $SU(N-5)$ to $SU(N-5-r)$.
The duality is clearly preserved when $r\le N-8$.  
If $r=N-7$, the magnetic gauge group is $SU(2)$, the
field $s$ is massive and $W=M^6Y^2 +M^2 (q^2)^2 +Y q q'$
(as noted above, the dynamical source of the $M^6Y^2$ term
is unknown.)  If $r\ge N-6$, the $SU(2)$ group is spontaneously broken.
At this stage the electric theory confines.  (In some regions of moduli
space the electric gauge group is broken and the magnetic theory confines; 
but there is no invariant distinction between the confining and Higgs
phases in this theory.)
We can keep on studying the dual which is a theory of singlets.
Alternatively, we can study the electric theory 
 directly using the techniques of~\nativacua. 

First consider $r=N-6$, leaving six massless vectors.
Both analyses yield the result
that the light fields are the independent gauge
singlets $M_{ij}, Y_i, B^i$ (recall $B^i$ is 
$\epsilon^{ij_1\dots j_5}Q_{j_1}\cdots Q_{j_5}P^2$ and $q^i$
in the electric and magnetic theories respectively)
satisfying the two constraints 
\eqn\constraints{\ev{Y_iB^i}=0 \qquad {\rm and}
\qquad \ev{M^5Y^2+MB^2}=C}
 on their expectation values, 
where $C$ is a constant. 
The 't Hooft anomaly matching conditions are
satisfied. In the classical limit
$\Lambda_{electric}\to 0$, we have $C\to 0$ and 
these constraints follow {}from $Spin(10)$
Fierz identities.
We may further check these constraints in the following two ways.
Giving an expectation value 
$\vev{M}$ of rank 6 for the fields $M$ that
have not been integrated out, we find that the
gauge group is broken to $Spin(4)\approx SU(2)_L\times SU(2)_R$; each
gauge group has four doublets $P^a_S$ ($a=1,2,3,4;S=L,R$),
coming {}from the spinor $P$, whose mesons 
$V^{ab}_S=P^a_SP^b_S$ satisfy
the {\sl two} constraints
 $\Pf \ V_S=\Lambda_S^4$ ($S=L,R$) \nativacua. 
The operators $V$ are linear combinations of $Y$ and $B$
and the two $Spin(10)$ constraints flow
to linear combinations of the $Spin(4)$ constraints.
Similarly, along an $SU(3)$ flat direction of the $Spin(10)$ theory,
where the massless field $\vev{M}$ has rank 2 and $\vev{Y}\neq 0$, 
three flavors of ${\bf 3}$, ${\bf \overline{3}}$ remain.
This theory satisfies only {\sl one}
 constraint \nativacua.  The two constraints
of the $Spin(10)$ theory relate a reducible $SU(3)$ operator to 
irreducible ones, and can be combined into a single constraint on 
mesons and baryons of $SU(3)$.

For $r=N-5$, 
a superpotential proportional to $1/(M^4Y^2+B^2)$
is obtained. It is generated by instantons in the electric theory. 
When $N-r=1,2,3,4$ vectors remain massless,
gaugino condensation in an unbroken
subgroup of $Spin(10)$ generates a superpotential 
proportional to $\left[1/M^{N-r-1}Y^2\right]^{1/(6-N+r)}$.
Clearly the equation of motion for $Y$ requires supersymmetric
solutions to be at infinite values of the fields.

\newsec{Dynamical SUSY breaking}

We now wish to study supersymmetry breaking.
As mentioned in the introduction, it is believed that $Spin(10)$ with 
one {\bf 16} and no {\bf 10}s
breaks SUSY dynamically\foot{We recall that
a solution of the 't Hooft anomaly matching 
conditions for unbroken
$U(1)_R$ was found in \adsspin, but was 
discarded because of its complexity. We see no trace 
of such a solution in our analysis
of the dual theory.} \adsspin. This theory 
has no chiral gauge-invariant operators and
thus no flat directions; it is therefore very strongly coupled. 
Recently, Murayama \mur\ studied the
theory with $N=1$. Classically there is a moduli space, which
allowed him to study the theory at weak coupling.
Under mass and Yukawa coupling perturbations, 
he showed that SUSY is broken at weak coupling
in the Higgs phase.  As the mass for the vector goes to infinity, 
the SUSY-breaking vacuum moves to the origin. There the theory is
strongly coupled and theoretical control is lost. 

Here, we present new evidence that SUSY is dynamically broken.
We study the strong coupling
regime of the theory of one spinor and $N$ massive
vectors, using the weakly coupled magnetic description. 
The dual provides the correct degrees of freedom near the origin
of the moduli space: the fields $M$, $Y$, $q$, etc.
We can look for a supersymmetric vacuum by
seeking solutions of the equations of motion for these fields.
When small masses are given to all $N$ vectors, these 
equations have no solution.  By holomorphy, we may continue
these masses to infinite values without restoring 
supersymmetry\foot{Supersymmetry can be
restored only on surfaces of {\sl complex codimension one} 
in parameter space, so there exist curves along which all masses may be
taken to infinity.}, implying that SUSY is broken in 
the $Spin(10)$ theory with one spinor and no vectors. 

We will concentrate on the case $N=6$. At this value of $N$ we
have a whole moduli space of vacua where the 't Hooft anomaly
matching conditions are satisfied, but no magnetic gauge dynamics which
could introduce non-perturbative phenomena.  The cases $1\le N\le 5$, 
where the only vacua are at infinite field expectation values, can be
derived {}from the case $N=6$ (using holomorphy.)
For $N\ge 7$, we may (by holomorphy) take all but six vectors
to be very massive; the Higgs mechanism breaks the magnetic gauge group
leaving the theory of singlets dual to the case $N=6$.

For $N=6$, the singlet fields
satisfy the two constraints \constraints, 
which we implement in the superpotential
using Lagrange multipliers $X_1, X_2$. 
Now add to the superpotential $mM+yY$; SUSY
should be broken for generic values of the coupling $y$ 
and of the rank six matrix $m$.
We claim the equations of motion for the magnetic superpotential 
$W = mM+yY+X_1(M^5Y^2 + MB^2 - C)+X_2YB$
have no solutions in its region of validity (near the origin of
moduli space, at strong coupling.)  {}From $M_{ij}(\del W/\del M_{ij})=0$
we learn finite $\vev{M}$ implies finite $X_1$.
Noting  that $(M^4Y^2)^{ij}Y_j=0$ by 
antisymmetry, and using  $Y_{j}(\del W/\del M_{ij})=0$ and $Y_jB^j=0$,
we find $m^{ij}Y_j=0$; since $m$ is rank six, $Y=0$. 
But $B^iB^j$ has rank at most one, so $\det[\del W/\del M]=\det m = 0$,
contrary to assumption.
This phenomenon is to be interpreted as dynamical supersymmetry 
breaking in the strongly coupled electric theory and as 
tree level, O'Raifeartaigh-type breaking in the
infrared-free magnetic theory of singlets.

One may use other duality 
transformations to produce more theories that
break SUSY.  Consider the magnetic theory for $N=7$,
which is $SU(2)\approx Sp(1)$ with eight doublets $q$ and $q'$ and
with $W = M^6Y^2 + M^2(q^2)^2 + 
Yqq'+mM$. We now dualize it according to the $Sp$ duality of \sem.   
We obtain a dual representation which is $Sp(1)$ with eight doublets
$d$ and $d'$, and gauge singlets $M$, $Y$, $\hat B=q^2$ and 
$\tilde Y=qq'$.  Integrating out the massive fields $Y,\tilde Y$ 
leaves the superpotential 
$W = M^6 (dd')^2+ M^2\hat B^2 + \hat B dd + mM$. As expected, 
this theory breaks SUSY when $m$ has maximal rank 7.
To see that, first note that if $m$ has rank 1, the field $\hat B$ may
have rank 2, giving mass to two of the doublets $d$, leaving
six doublets and causing the theory to confine \nativacua. 
(In other regions of moduli space the $Sp(1)$ gauge group is
broken, but there is no distinction between confining
and Higgs phases in this theory.) 
The spectrum and contraints of the resulting theory of singlets
are then identical to the $N=6$ theory studied in the previous 
paragraph, and the analysis for $m$ of higher rank follows immediately.

As another possibility, we can add two singlets $X, \tilde X$ 
(antisymmetric tensors of $SU(2)$!) to the magnetic $SU(2)$
theory, with a potential $(X\tilde X)^{k}$; this does
not affect the dynamics.  This theory has
a dual description \ilstr\ as $SU(4k+2)$ with a flavor 
of fields $x,\tilde x$
in the antisymmetric tensor representation,
four flavors of fundamental and antifundamental representations,
numerous singlets and a complicated superpotential.  The specific 
dynamics of the resulting SUSY breaking depends on $k$.
Similar tricks may be used to create huge classes of theories that
break supersymmetry.  Ideas of this type have been illustrated
in \ithomas.

It would be interesting to exhibit duality in a theory that flows down
to the $SU(2)$ model with matter in the four-dimensional (spin $3/2$)
representation which was 
argued in \iss\ to break SUSY dynamically.

\centerline{{\bf Acknowledgments}}

We would like to thank K. Intriligator, R. Leigh, A. Nelson, N. Seiberg
and S. Thomas for useful discussions.
 This work was supported in part by DOE grant \#DE-FG05-90ER40559
and by a Canadian 1967 Science fellowship.
\listrefs
\bye